\documentclass[a4paper]{article}
\usepackage{latexsym}
\usepackage{mymacros}
\usepackage{axodraw}
\title{The Electroweak Standard Model in the Axial Gauge}
\author{Chris Dams\footnote{chrisd@sci.kun.nl}\hspace{1cm}
	Ronald Kleiss\footnote{kleiss@sci.kun.nl}\\
	\normalsize University of Nijmegen\\
	\normalsize	Institute for Theoretical Physics\\
	\normalsize Toernooiveld~1\\
	\normalsize 6525 ED~~Nijmegen\\
	\normalsize The Netherlands}
\date{March 11, 2004}
\def\beq{\begin{equation}}
\def\eeq{\end{equation}}
\def\textscr#1{\textrm{\scriptsize #1}}
\def\tfrac#1#2{\textstyle\frac{#1}{#2}}
\let\oslash\slash
\def\slash#1{\setbox0\hbox{$#1$}\hbox to\wd0{\hss$/$\hss}\nobreak\hskip-\wd0\box0}
\def\fmruletab#1{{\openup1ex\halign{\hskip 1cm\hfil$\vcenter{\hsize=50pt\noindent##}$\hfil &\hskip 1cm $\displaystyle##$\hfil\cr#1}}}
\begin{document}
\def\L{\mathcal L}
\maketitle

\begin{abstract}
\noindent We derive the Feynman rules of the standard model in the axial
gauge. After this we prove that the fields~$\phi_W$ and $\phi_Z$
do not correspond to physical particles. As a consequence, these
fields cannot appear as incoming or outgoing lines in Feynman
graphs. We then calculate the contribution of these fields in the case
of a particular decay mode of the top quark.
\end{abstract}

\section{Introduction}
We consider the electroweak standard model in the axial gauge, restricting
ourselves to leptons for simplicity. We include Dirac
masses for the neutrino's, not just because these particles appear to have
a mass, but mainly
to make it easier to figure out what the Feynman rules for the quarks are.
The reason to consider the standard model in this gauge is
that it can provide a more severe check on gauge invariance than the more
common gauges.  In~\cite{gauge} an example of a gauge dependent quantity was
found that in the $R_\xi$-gauge did not depend on the
gauge parameter~$\xi$ but in the axial gauge did depend on
the gauge vector~$n$. Another advantage of this gauge is that no
Fadeev-Popov ghost particles are needed. There are, however, unphysical
bosonic particles. Both kinds of unphysical particles disappear in tree
graphs in the unitary gauge, but reappear in loop graphs.
Furthermore, the unitary gauge has no gauge parameter, so the only practical
check on the gauge invariance of a cross section is its high energy behaviour.
The disadvantage of the axial gauge is that one either has bilinear terms
in unphysical bosonic degrees of freedom and $W$ or $Z$ particles or, if one
diagonalizes these, rather complicated formulae for interaction vertices (and
in addition quite a lot of different interaction vertices).
We choose the option of having diagonalized propagators.

\section{The Lagrangian}
Many lecture notes and books contain introductions to the standard model,
see for instance, \cite{bsm}. Here, we just quickly recall the terms of the
Lagrangian of the unbroken standard model. After that we turn to the
axial gauge. The electroweak standard model has
$\textrm{SU}(2)\times\textrm U(1)$ as its gauge group.
The gauge field that belongs to
$\textrm{SU}(2)$ is called~$A^a_\mu$, with $a=1,2,3$. The gauge field that
belongs to $\textrm U(1)$ is called~$B_\mu$.
The lefthanded fermions are in the $(2,-\frac12)$ representation and the 
righthanded ones are in $(1,-1)$. Furthermore there are righthanded neutrino's
in the trivial representation of the gauge group. This means that the
Lagrangian for the fermions is
\beq
\L_{\textrm{\scriptsize fermion}}
   =\bar\psi_L(i\slash\partial-g_2\slash A^a T^a
            +\tfrac12g_1\slash B)\psi_L
   +\bar\psi_R(i\slash\partial+g_1\slash B)\psi_R
	+\bar\psi_\nu (i\slash\partial)\psi_\nu,
\eeq
where $\psi_\nu$ stands for the right-handed neutrino field.
Note that the $T^a$ are $2\times2$-matrices that act on the two components
of~$\psi_L$. It looks as if the $\psi_\nu$~field is not coupled to anything
but that will change if we introduce the field~$\phi$ below. The Lagrangian
for the gauge fields is
\beq
\eqalign{
\L_{\textrm{\scriptsize gauge}}&=
		-\tfrac12(\partial^\nu B^\mu)(\partial_\nu B_\mu)
      +\tfrac12(\partial^\mu B_\mu)(\partial^\nu B_\nu)
      -\tfrac12(\partial^\nu A^{a\mu})(\partial_\nu A^a_\mu)\cr
      &\qquad+\tfrac12(\partial^\mu A^a_\mu)(\partial^\nu A^a_\nu)
      +g_2\epsilon^{abc}(\partial^\mu A^{a\nu})A^b_\mu A^c_\nu\cr
      &\qquad-\tfrac14g_2^2A^{a\mu}A^a_\mu A^{b\nu}A^b_\nu
      +\tfrac14g_2^2A^{a\mu}A^b_\mu A^{a\nu}A^b_\nu.\cr
}
\eeq
Furthermore there is a complex scalar field~$\phi$ in the
$(2,\frac12)$ representation. This has the Lagrangian
\beq
\eqalign{
\L_{\textrm{\scriptsize scalar}}&=
   (\partial_\mu\phi)^\dagger(\partial^\mu\phi)
   +ig_2A^{a\mu}(\partial_\mu\phi)^\dagger T^a\phi
   -ig_2A^{a\mu}\phi^\dagger T^a(\partial_\mu\phi)\cr
   &\qquad+\tfrac14g_2^2A^a_\mu A^{a\mu}\phi^\dagger\phi
   +\tfrac{ig_1}2B^\mu(\partial_\mu\phi)^\dagger\phi
   -\tfrac{ig_1}2B^\mu\phi^\dagger(\partial_\mu\phi)\cr
   &\qquad+g_1g_2A^a_\mu B^\mu\phi^\dagger T^a\phi
   +\tfrac{g_1^2}4B^2\phi^\dagger\phi
	-\mu^2\phi^\dagger\phi-\tfrac{\lambda_\phi}4(\phi^\dagger\phi)^2.\cr
}
\eeq
Finally, we can couple the field~$\phi$ to the fermions. The Lagrangian
is called the Yukawa Lagrangian. It is given by
\beq
\L_{\textrm{\scriptsize Yukawa}}
   =g_{\alpha\beta}\bar\psi^\alpha_L\phi\psi_R^\beta
   +g_{\alpha\beta}^\dagger\bar\psi_R^\alpha\phi^\dagger\psi_L^\beta
	+h_{\alpha\beta}\bar\psi_L^\alpha\epsilon\phi^*\psi_\nu^\beta
	-h^\dagger_{\alpha\beta}\bar\psi_\nu^\alpha\phi^T\epsilon\psi_L^\beta.
\eeq
The indices $\alpha$~and~$\beta$ enumerate the generations of the standard
model and the matrices $g$ and $h$ contain complex numbers that can, in
principle, be chosen freely. $\epsilon$ is the two-dimensional Levi-Civita
tensor. It is not difficult to see that all these terms transform
trivially under the gauge group. The reason that it is possible to construct
an $\mathrm{SU}(2)$ invariant from $\psi_L$ and $\phi$ as well as from
$\bar\psi_L$ and $\phi$ is that the fundamental representation of
$\mathrm{SU}(2)$ is pseudo-real. The reality of representations is, for
instance, discussed in~\cite{group}.

We briefly outline the symmetry breaking using the axial gauge fixing.
The unbroken standard model, as defined by the above Lagrangians,
is invariant under
local gauge transformations. The fermion fields and the field~$\phi$ transform
according to the representation they are in.
The vector fields transform according to the infinitesimal transformations
\beq
\eqalign{
\delta B_\mu&=\tfrac1{g_1}(\partial_\mu(\delta\Lambda));\cr
\delta A^a_\mu
	&=\tfrac1{g_2}(\partial_\mu(\delta\Lambda^a))
											+\epsilon^{abc}(\delta\Lambda^b)A_\mu^c,\cr
}
\eeq
if we parameterize group elements by~$e^{-i\Lambda}$ and $e^{-i\Lambda^aT^a}$.
$\Lambda$ and $\Lambda^a$ are four arbitrary functions of space-time. The
freedom to choose four arbitrary functions of space-time indicates that
there is a large redundancy in the field configurations. In the path
integral this redundancy causes problems, because of integrating over many
equivalent field configurations, and we need to get rid of it.
The various ways of doing this are the various gauges. We choose the
so-called axial gauge. This means that we add to the Lagrangian the
quantity
\beq
\L_{\textscr{gauge-fixing}}=
	-\tfrac12\lambda n^\mu A^a_\mu A^a_\nu n^\nu
   -\tfrac12\lambda (n\cdot B)^2,
\eeq
and in the resulting Feynman rules take the limit~$\lambda\to\infty$.
The various gauges should give the same observable results (e.g. cross
sections) and these should not depend on parameters in the gauge choice.
In our case they should not depend on the gauge vector~$n$.

In the standard model, it is assumed that the parameter~$\mu^2$ that
appears in the Lagrangian for the scalar field~$\phi$ is negative.
The consequence of this is that the minimum of the energy of this field is no
longer located at the point~$\phi=0$, but instead at the sphere
$\phi^\dagger\phi=-2\mu^2/\lambda_\phi$. To derive Feynman rules, we make
the substitution
\beq
\phi\to\frac1{\sqrt 2}\pmatrix{0\cr v\cr}+\phi,
\eeq
with~$v=2\sqrt{-\mu^2/\lambda_\phi}$ so that the potential is minimal
for~$\phi=0$. The different components of the $\phi$-field get different
r\^oles because of the arbitrary choice of the direction of the translation
of the $\phi$-field. The second component of this complex field is split
into two real components according to
\beq
\phi_2=\frac1{\sqrt2}\left(H+i\phi_Z\right).
\eeq
After the field translation, the fields $A^3$~and~$B$ mix
in the bilinear terms. We have
\beq
\eqalign{
\L_{A^3B,\textscr{bilinear}}&=
		-\tfrac12(\partial^\nu A^{3\mu})(\partial_\nu A^3_\mu)
      +\tfrac12(\partial^\mu A^3_\mu)(\partial^\nu A^3_\nu)
		+\tfrac18g_2^2v^2 A^3_\mu A^{3\mu}\cr
		&\qquad-\tfrac12\lambda n^\mu A^a_\mu A^a_\nu n^\nu
		-\tfrac12(\partial^\nu B^\mu)(\partial_\nu B_\mu)
      +\tfrac12(\partial^\mu B_\mu)(\partial^\nu B_\nu)\cr
		&\qquad+\tfrac18g_1^2v^2 B_\mu B^\mu
   	-\tfrac12\lambda (n\cdot B)^2-\tfrac14g_1g_2v^2A^3_\mu B^\mu.\cr
}
\eeq
This part of the Lagrangian can be diagonalized by making the substitution
\beq
\eqalign{
	A^3_\mu&\to\cos\theta_wA^3_\mu+\sin\theta_wB_\mu;\cr
	B_\mu&\to\cos\theta_wB_\mu-\sin\theta_wA^3_\mu,\cr
}
\eeq
with $\cos\theta_w=g_e/g_1$ and $\sin\theta_w=g_e/g_2$. $g_e$ is by definition
given by $g_e^2=g_1^2g_2^2/(g_1^2+g_2^2)$. At this point, we
introduce the masses $M_H$~and~$M_W$. These are given by
\beq
\eqalign{
M_W&=\frac{g_ev}{2\sin\theta_w};\cr
M_H^2&=\tfrac12\lambda_\phi v^2.\cr
}
\eeq
The field~$H$ turns out to have a
mass~$M_H$, while $M_W$ is the mass of the fields~$A^{1,2}$. The field~$B$
has become massless, the mixing term between $A^3$ and $B$ has disappeared,
and the field $A^3$ has gotten a mass~$M_Z=M_W/\cos\theta_w$. At this
point we change the name of the field~$A^3$ into $Z$, and the components
$A^{1,2}$ are taken to be the real and imaginary parts of the complex
vector field~$W$ according to
\beq
\label{eq:Wdef}
\eqalign{
A^1_\mu&=\frac1{\sqrt2}\left(W^{\phantom*}_\mu+W^*_\mu\right);\cr
A^2_\mu&=\frac1{i\sqrt2}\left(W^{\phantom*}_\mu-W^*_\mu\right).\cr
}
\eeq

We still have a mixing term between $\phi_Z$~and~$Z$. The bilinear
terms in these fields are given by
\beq
\eqalign{
\L_{Z\phi_Z,\textscr{bilinear}}&=
		-\tfrac12(\partial^\nu Z^\mu)(\partial_\nu Z_\mu)
      +\tfrac12(\partial^\mu Z_\mu)(\partial^\nu Z_\nu)
		+\tfrac12M_Z^2Z_\mu Z^\mu\cr
		&\qquad-\tfrac12\lambda n^\mu Z_\mu Z_\nu n^\nu
		+\tfrac12(\partial^\mu\phi_Z)(\partial_\mu\phi_Z)
		-M_ZZ^\mu\partial_\mu\phi_Z.
}
\eeq
This part of the Lagrangian can be diagonalized in momentum space by
substituting
\beq
\phi_Z(k)\to\phi_Z(k)+2iM_Z\frac{k^\mu Z_\mu(k)}{k^2}.
\eeq
It is inadvisable to make a substitution on~$Z$, because of the presence
of the gauge vector~$n$.

For the fields $W$~and~$\phi_1$ (i.e., the first component
of the complex $\phi$-field) we have a situation similar to what we
had for the fields $Z$~and~$\phi_Z$. These fields still mix. We have in the
Lagrangian the bilinear terms
\beq
\eqalign{
\L_{W\phi,\textscr{bilinear}}&=-(\partial^\mu W^\nu)(\partial_\mu W^*_\nu)
	+(\partial^\mu W_\mu)(\partial^\nu W^*_\nu)
	+M_W^2 W^\mu W^*_\mu\cr
	&\qquad-\lambda n^\mu W_\mu n^\nu W^*_\nu
	+(\partial^\mu \phi_1)(\partial_\mu\phi_1^*)\cr
	&\qquad+iM_W W^*_\mu\partial^\mu\phi_1^*
	-iM_W W^\mu\partial^\mu\phi_1.\cr
}
\eeq
The reason that, in the mixing terms, we have two conjugated fields or 
two unconjugated fields is because of the way that we chose to put the
fields $A^{1,2}$ into the complex field~$W$ in equation~(\ref{eq:Wdef}).
We chose this way, because
it gives the normal conventions in the couplings to the fermions.
These terms are diagonalized by applying, in momentum space, the
transformation
\beq
\phi_1(k)\to\phi_1(k)+2M_W \frac{k^\mu W^*_\mu(-k)}{k^2}.
\eeq
For convenience, we rename the field $\phi_1$ into $\phi_W^*$ and
$\phi_1^*$ into $\phi_W$.

After the diagonalization process the quadratic terms in the Lagrangian
for the field~$Z$ are, in momentum space, given by
\beq
\eqalign{
\L_{Z^2}&=-\tfrac12 k^2 Z(k)^\mu Z(-k)_\mu
	+\tfrac12k^\mu Z(k)_\mu k^\nu Z(-k)_\nu
	+\tfrac12M_Z^2 Z(k)^\mu Z(-k)_\mu\cr
	&\qquad-\tfrac12\tfrac{M_Z^2}{k^2}k^\mu Z(k)_\mu k^\nu Z(-k)_\nu
	-\tfrac12\lambda n^\mu Z(k)_\mu n^\nu Z(-k)_\nu.\cr
}
\eeq
From this the propagator
\beq
\Delta_{\nu\mu}=\frac{-i\left(
      g_{\nu\mu}
      -\frac{n_\nu k_\mu+n_\mu k_\nu}{n\cdot k}
      +k_\nu k_\mu\frac{n^2+(k^2-M_Z^2)/\lambda}{(n\cdot k)^2}
   \right)}{k^2-M_Z^2+i\epsilon}
\eeq
can be found. Taking the limit $\lambda\to\infty$, the term with
$(k^2-M_Z^2)/\lambda$ disappears and we see that the numerator is
the same as in the axial gauge for massless particles. In the rest
of this paper this limit is implied. For the $W$-particle the same
propagator can be found except that $M_Z$ should be changed
into~$M_W$.

From this propagator we can derive the polarization sum in the axial
gauge. In the theory without interaction we have a particle creation
field configuration
\beq
a_j^*(\vec k)=-i\int d^3x\,
	e^{-ik\cdot x}
	\mathop{\raise 1pt\hbox{$\mathop\partial\limits^\leftrightarrow$}}\nolimits_0
	s_j^\mu Z_\mu(x).
\eeq
The complex conjugate of this is the particle annihilation field
configuration.
Because the vector field~$Z$ has three physical degrees of freedom,
the $j$ in the above formula should run from 1~to~3. For
$s_{1,2}^\mu$ we choose two vectors perpendicular to each other and
perpendicular to both $k$ and $n$ with $s_{1,2}^2=-1$. For $s_3$
we pick
\beq
s_3^\mu=\frac{n\cdot k}{M_Z\sqrt{(k\cdot n)^2-k^2n^2}}k^\mu.
\eeq
This is correctly normalized as can be checked by verifying that
\beq
\langle a_3^*(\vec k,t)a_3(\vec k',t')\rangle
	=2\sqrt{|\vec k|^2+M_Z^2}\,(2\pi)^3\delta^3(\vec k-\vec k')\theta(t'-t).
\eeq
It is possible to add an arbitrary multiple of $n^\mu$ in the
definition of $s_3^\mu$, but
since the contraction of $n^\mu$ with the propagator is zero, this
does not contribute. The polarization vectors~$\epsilon_j$
that occur in the Feynman rules are the contraction of $s_j$ with the
numerator of the propagator. We have
\beq
\epsilon_{j\mu}=-\left(g_{\mu\nu}
		-\frac{n_\nu k_\mu}{n\cdot k}
		-\frac{n_\mu k_\nu}{n\cdot k}
      +k_\nu k_\mu\frac{n^2}{(n\cdot k)^2}\right)s_j^\nu.
\eeq
From this it can be found that the polarization sum is given by
\beq
\sum_{j=1,2,3}\epsilon_j^\mu\epsilon_j^\nu
	=-g^{\mu\nu}
      +\frac{n^\nu k^\mu}{n\cdot k}
		+\frac{n^\mu k^\nu}{n\cdot k}
      -k^\nu k^\mu\frac{n^2}{(n\cdot k)^2}.
\eeq
In practice, only the $-g^{\mu\nu}$ term plays a role, because
it is a feature of the axial gauge that if we have a vector boson
($B$,~$W$ or~$Z$) as an incoming/outgoing particle, the matrix element
should become zero if a polarization
vector is replaced by the momentum of the external particle the
polarization vector belongs to. This is a check on gauge invariance.
Note that it is an error to contract the polarization sum with the
numerator of the propagator. In the axial gauge one should be careful
not to confuse the vectors~$s^\mu$ with the vectors~$\epsilon^\mu$.

Also the fermions can be diagonalized. This proceeds in exactly the same
way as in more common gauges. The result is that there are six different
fermion masses and that the coupling to the $W$~boson can change a fermion
of one generation into a fermion of another.

\newpage
\section{Feynman Rules}
Below we list the Feynman rules of the standard model in the axial gauge.
A few remarks are in order
\begin{enumerate}
\item For every Feynman rule that involves fermions, there is another one
	with all generation labels changed. This involves the changes
	$e\leftrightarrow \mu$, $\nu_e\leftrightarrow\nu_\mu$,
	$m_e\leftrightarrow m_\mu$ and
	$m_{\nu_e}\leftrightarrow m_{\nu_\mu}$. Furthermore, in subscripts of the
	neutrino mixing matrix~$V$ the exchange $1\leftrightarrow 2$ should be
	carried out. Also one of the generations involved can be changed into
	the third generation (i.e., the $\tau$~fermion). Of Feynman rules related
	in this way, only one is shown below
\item Particles that have an antiparticle, have an arrow on their lines
	in a Feynman graph. In this case, momentum flows in the direction of
	the arrow. If particles do not have an arrow on them, momentum flows
	towards the vertex.
\item We use the following abbreviations
	\beq
	\eqalign{
	g_w&=\frac{g_e}{\sin\theta_w};\cr
	g_z&=\frac{g_e}{\sin\theta_w\cos\theta_w};\cr
	p_l&=\tfrac12(1-\gamma^5);\cr
	p_r&=\tfrac12(1+\gamma^5).\cr
	}
\eeq
\item If reversing all arrows on a vertex would yield a different vertex,
	that vertex is also a vertex of the theory. To find the vertex factor
	that belongs to it, the vertex factor of the original vertex should
	be complex conjugated, except for one factor of~$i$, and all momenta
	that belong to particles that do not carry an arrow on their line
	should get a minus sign. Of a pair of vertices that is related in
	this way, only one is shown below.
	As an example, consider the vertex
	with an incoming electron neutrino, an outgoing muon and an incoming
	$\phi_W$, that is shown below. The ``conjugate vertex factor'' is found by
	exchanging $p_r$ and~$p_l$ and changing $V^\dagger_{21}$ into~$V_{12}$. 
	Another example is the vertex with an incoming Higgs, an
	incoming~$\phi_W$ and an outgoing~$W$ (see below). To obtain the
	vertex that belongs to an incoming Higgs, an incoming~$W$ and an
	outgoing~$\phi_W$, the only change necessary in the vertex factor is
	$k_1\to-k_1$.
\item The algebra necessary to find all the vertex factors was done using
	the C++ computer algebra library GiNaC, see~\cite{ginac}. Because
	other symbolic calculations will be easier to perform starting from
	the Lagrangian calculated here, the program used can be downloaded
	at the homepage~\cite{home} of one of the authors.
\end{enumerate}

\newpage
\subsection{Propagators}
\fmruletab{
\begin{picture}(50,50)(0,0)
\Line(0,25)(50,25)\Text(25,26)[b]{$\scriptscriptstyle B(k)$}
\end{picture}
&
\frac{-i\left(
      g_{\nu\mu}
      -\frac{n_\nu k_\mu+n_\mu k_\nu}{n\cdot k}
      +k_\nu k_\mu\frac{n^2}{(n\cdot k)^2}
   \right)}{k^2+i\epsilon}
\cr
\begin{picture}(50,50)(0,0)
\ArrowLine(0,25)(50,25)\Text(25,26)[b]{$\scriptscriptstyle W(k)$}
\end{picture}
&
\frac{-i\left(
      g_{\nu\mu}
      -\frac{n_\nu k_\mu+n_\mu k_\nu}{n\cdot k}
      +k_\nu k_\mu\frac{n^2}{(n\cdot k)^2}
   \right)}{k^2-M_W^2+i\epsilon}
\cr
\begin{picture}(50,50)(0,0)
\ArrowLine(0,25)(50,25)\Text(25,27)[b]{$\scriptscriptstyle \phi_W(k)$}
\end{picture}
&
\frac i{k^2}
\cr
\begin{picture}(50,50)(0,0)
\Line(0,25)(50,25)\Text(25,26)[b]{$\scriptscriptstyle Z(k)$}
\end{picture}
&
\frac{-i\left(
      g_{\nu\mu}
      -\frac{n_\nu k_\mu+n_\mu k_\nu}{n\cdot k}
      +k_\nu k_\mu\frac{n^2}{(n\cdot k)^2}
   \right)}{k^2-M_Z^2+i\epsilon}
\cr
\begin{picture}(50,50)(0,0)
\Line(0,25)(50,25)\Text(25,26)[b]{$\scriptscriptstyle \phi_Z(k)$}
\end{picture}
&
\frac i{k^2}
\cr
\begin{picture}(50,50)(0,0)
\Line(0,25)(50,25)\Text(25,26)[b]{$\scriptscriptstyle H(k)$}
\end{picture}
&
\frac i{k^2-M_H^2+i\epsilon}
\cr
\begin{picture}(50,50)(0,0)
\ArrowLine(0,25)(50,25)\Text(25,27)[b]{$\scriptscriptstyle e(k)$}
\end{picture}
&
\frac{i(\slash k+m_e)}{k^2-m_e^2+i\epsilon}
\cr
\begin{picture}(50,50)(0,0)
\ArrowLine(0,25)(50,25)\Text(25,27)[b]{$\scriptscriptstyle \nu_e(k)$}
\end{picture}
&
\frac{i(\slash k+m_{\nu_e})}{k^2-m_{\nu_e}^2+i\epsilon}
\cr
}

\subsection{Triple boson couplings without Higgs}
\fmruletab{
\begin{picture}(50,50)(0,0)
\ArrowLine(0,0)(19,25)\Text(10,12)[lt]{$\scriptscriptstyle W(k_1)^\mu$}
\ArrowLine(19,25)(0,50)\Text(10,38)[lb]{$\scriptscriptstyle W(k_2)^\nu$}
\Line(19,25)(50,25)\Text(34,26)[b]{$\scriptscriptstyle B(k_3)^\sigma$}
\end{picture}
&
\eqalign{
ig_e\left[\vphantom{\frac12}\right.&
		g^{\nu\sigma}(k_2^\mu+k_3^\mu)
		+g^{\mu\sigma}(k_1^\nu-k_3^\nu)
		-g^{\mu\nu}(k_1^\sigma+k_2^\sigma)\cr
		&\left.-M_W^2\left(
			g^{\nu\sigma}\frac{k_1^\mu}{k_1^2}
			+g^{\mu\sigma}\frac{k_2^\nu}{k_2^2}
			-(k_1^\sigma+k_2^\sigma)\frac{k_1^\mu}{k_1^2}\frac{k_2^\nu}{k_2^2}
		\right)
	\right]
}
\cr
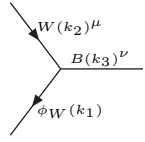
\begin{picture}(50,50)(0,0)
\ArrowLine(19,25)(0,0)\Text(10,12)[lt]{$\scriptscriptstyle \phi_W(k_1)$}
\ArrowLine(0,50)(19,25)\Text(10,38)[lb]{$\scriptscriptstyle W(k_2)^\mu$}
\Line(19,25)(50,25)\Text(34,26)[b]{$\scriptscriptstyle B(k_3)^\nu$}
\end{picture}
&
ig_eM_W\left(g^{\mu\nu}-\frac{(k_1^\nu+k_2^\nu)k_2^\mu}{k_2^2}\right)
\cr
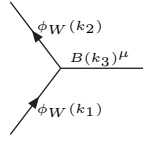
\begin{picture}(50,50)(0,0)
\ArrowLine(0,0)(19,25)\Text(10,12)[lt]{$\scriptscriptstyle \phi_W(k_1)$}
\ArrowLine(19,25)(0,50)\Text(10,38)[lb]{$\scriptscriptstyle \phi_W(k_2)$}
\Line(19,25)(50,25)\Text(34,26)[b]{$\scriptscriptstyle B(k_3)^\mu$}
\end{picture}
&
ig_e(k_1^\mu+k_2^\mu)
\cr
\begin{picture}(50,50)(0,0)
\ArrowLine(0,0)(19,25)\Text(10,12)[lt]{$\scriptscriptstyle W(k_2)^\nu$}
\ArrowLine(19,25)(0,50)\Text(10,38)[lb]{$\scriptscriptstyle W(k_3)^\sigma$}
\Line(19,25)(50,25)\Text(34,26)[b]{$\scriptscriptstyle Z(k_1)^\mu$}
\end{picture}
&	\eqalign{
	ig_w\cos\theta_w\left[\vphantom{\frac12}\right.&
	-g^{\nu\sigma}(k_2^\mu+k_3^\mu)
	-g^{\mu\nu}(k_1^\sigma-k_2^\sigma)
	+g^{\mu\sigma}(k_1^\nu+k_3^\nu)\cr
	&+M_Z^2\sin^2\theta_w\left(
		g^{\mu\sigma}\frac{k_2^\nu}{k_2^2}
		+g^{\mu\nu}\frac{k_3^\sigma}{k_3^2}
	\right)\cr
	&+\frac12M_Z^2\left(
		-(k_1^\sigma-k_2^\sigma)\frac{k_1^\mu}{k_1^2}\frac{k_2^\nu}{k_2^2}
		-(k_1^\nu+k_3^\nu)\frac{k_1^\mu}{k_1^2}\frac{k_3^\sigma}{k_3^2}
	\right)\cr
	&\left.+M_Z^2\left(\frac12-\sin^2\theta_w\right)(k_2^\mu+k_3^\mu)
										\frac{k_2^\nu}{k_2^2}\frac{k_3^\sigma}{k_3^2}
	\right]\cr
	}
\cr
\begin{picture}(50,50)(0,0)
\ArrowLine(19,25)(0,0)\Text(10,12)[lt]{$\scriptscriptstyle \phi_W(k_2)$}
\ArrowLine(0,50)(19,25)\Text(10,38)[lb]{$\scriptscriptstyle W(k_3)^\nu$}
\Line(19,25)(50,25)\Text(34,26)[b]{$\scriptscriptstyle Z(k_1)^\mu$}
\end{picture}
&
-ig_zM_W\left(
	\sin^2\theta_wg^{\mu\nu}
	-\frac12\frac{(k_1^\nu+k_2^\nu)k_1^\mu}{k_1^2}
	+\left(\cos^2\theta_w-\frac12\right)\frac{(k_2^\mu+k_3^\mu)k_3^\nu}{k_3^2}
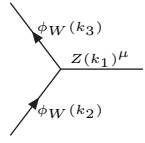
\right)
\cr
\begin{picture}(50,50)(0,0)
\ArrowLine(0,0)(19,25)\Text(10,12)[lt]{$\scriptscriptstyle \phi_W(k_2)$}
\ArrowLine(19,25)(0,50)\Text(10,38)[lb]{$\scriptscriptstyle \phi_W(k_3)$}
\Line(19,25)(50,25)\Text(34,26)[b]{$\scriptscriptstyle Z(k_1)^\mu$}
\end{picture}
&
ig_w\left(k_2^\mu+k_3^\mu\right)\left(\cos\theta_w-\frac1{2\cos\theta_w}\right)
\cr
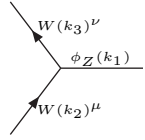
\begin{picture}(50,50)(0,0)
\ArrowLine(0,0)(19,25)\Text(10,12)[lt]{$\scriptscriptstyle W(k_2)^\mu$}
\ArrowLine(19,25)(0,50)\Text(10,38)[lb]{$\scriptscriptstyle W(k_3)^\nu$}
\Line(19,25)(50,25)\Text(34,26)[b]{$\scriptscriptstyle \phi_Z(k_1)$}
\end{picture}
&
\frac12g_wM_W\left(
	\frac{(k_2^\nu-k_1^\nu)k_2^\mu}{k_2^2}
	-\frac{(k_1^\mu+k_3^\mu)k_3^\nu}{k_3^2}
\right)
\cr
\begin{picture}(50,50)(0,0)
\ArrowLine(19,25)(0,0)\Text(10,12)[lt]{$\scriptscriptstyle \phi_W(k_2)$}
\ArrowLine(0,50)(19,25)\Text(10,38)[lb]{$\scriptscriptstyle W(k_3)^\mu$}
\Line(19,25)(50,25)\Text(34,26)[b]{$\scriptscriptstyle \phi_Z(k_1)$}
\end{picture}
&
\frac12g_w\left(k_1^\mu+k_2^\mu\right)
\cr
}

\subsection{Triple boson couplings with Higgs}
\fmruletab{
\begin{picture}(50,50)(0,0)
\ArrowLine(0,0)(19,25)\Text(10,12)[lt]{$\scriptscriptstyle W(k_2)^\mu$}
\ArrowLine(19,25)(0,50)\Text(10,38)[lb]{$\scriptscriptstyle W(k_3)^\nu$}
\Line(19,25)(50,25)\Text(34,26)[b]{$\scriptscriptstyle H(k_1)$}
\end{picture}
&
\frac i2g_wM_W\left(
	2g^{\mu\nu}
	-\frac{k_3^\nu(k_1^\mu+k_3^\mu)}{k_3^2}
	+\frac{k_2^\mu(k_1^\nu-k_2^\nu)}{k_2^2}
	-M_H^2\frac{k_2^\mu}{k_2^2}\frac{k_3^\nu}{k_3^2}
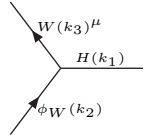
\right)
\cr
\begin{picture}(50,50)(0,0)
\ArrowLine(0,0)(19,25)\Text(10,12)[lt]{$\scriptscriptstyle \phi_W(k_2)$}
\ArrowLine(19,25)(0,50)\Text(10,38)[lb]{$\scriptscriptstyle W(k_3)^\mu$}
\Line(19,25)(50,25)\Text(34,26)[b]{$\scriptscriptstyle H(k_1)$}
\end{picture}
&
\frac i2g_w\left(k_2^\mu-k_1^\mu+\frac{M_H^2}{k_3^2}k_3^\mu\right)
\cr
\begin{picture}(50,50)(0,0)
\ArrowLine(0,0)(19,25)\Text(10,12)[lt]{$\scriptscriptstyle \phi_W(k_2)$}
\ArrowLine(19,25)(0,50)\Text(10,38)[lb]{$\scriptscriptstyle \phi_W(k_3)$}
\Line(19,25)(50,25)\Text(34,26)[b]{$\scriptscriptstyle H(k_1)$}
\end{picture}
&
-\frac i2g_w\frac{M_H^2}{M_W}
\cr
\begin{picture}(50,50)(0,0)
\Line(0,0)(19,25)\Text(10,12)[lt]{$\scriptscriptstyle Z(k_2)^\mu$}
\Line(19,25)(0,50)\Text(10,38)[lb]{$\scriptscriptstyle Z(k_3)^\nu$}
\Line(19,25)(50,25)\Text(34,26)[b]{$\scriptscriptstyle H(k_1)$}
\end{picture}
&
ig_zM_Z\left(
	g^{\mu\nu}
	+\frac12(k_1^\nu-k_2^\nu)\frac{k_2^\mu}{k_2^2}
	+\frac12(k_1^\mu-k_3^\mu)\frac{k_3^\nu}{k_3^2}
	+\frac12M_H^2\frac{k_2^\mu}{k_2^2}\frac{k_3^\nu}{k_3^2}
\right)
\cr
\begin{picture}(50,50)(0,0)
\Line(0,0)(19,25)\Text(10,12)[lt]{$\scriptscriptstyle \phi_Z(k_2)$}
\Line(19,25)(0,50)\Text(10,38)[lb]{$\scriptscriptstyle Z(k_3)^\mu$}
\Line(19,25)(50,25)\Text(34,26)[b]{$\scriptscriptstyle H(k_1)$}
\end{picture}
&
\frac12g_z\left(k_1^\mu-k_2^\mu+M_H^2\frac{k_3^\mu}{k_3^2}\right)
\cr
\begin{picture}(50,50)(0,0)
\Line(0,0)(19,25)\Text(10,12)[lt]{$\scriptscriptstyle \phi_Z(k_2)$}
\Line(19,25)(0,50)\Text(10,38)[lb]{$\scriptscriptstyle \phi_Z(k_3)$}
\Line(19,25)(50,25)\Text(34,26)[b]{$\scriptscriptstyle H(k_1)$}
\end{picture}
&
-\frac i2g_z\frac{M_H^2}{M_Z}
\cr
\begin{picture}(50,50)(0,0)
\Line(0,0)(19,25)\Text(10,12)[lt]{$\scriptscriptstyle H(k_2)$}
\Line(19,25)(0,50)\Text(10,38)[lb]{$\scriptscriptstyle H(k_3)$}
\Line(19,25)(50,25)\Text(34,26)[b]{$\scriptscriptstyle H(k_1)$}
\end{picture}
&
-\frac{3i}2g_w\frac{M_H^2}{M_W}
\cr
}

\subsection{Coupling to the Fermions}
\fmruletab{
\begin{picture}(50,50)(0,0)
\ArrowLine(0,0)(19,25)\Text(10,12)[lt]{$\scriptscriptstyle e(k_1)$}
\ArrowLine(19,25)(0,50)\Text(10,38)[lb]{$\scriptscriptstyle e(k_2)$}
\Line(19,25)(50,25)\Text(34,26)[b]{$\scriptscriptstyle B(k_3)^\mu$}
\end{picture}
&
ig_e\gamma^\mu
\cr
\begin{picture}(50,50)(0,0)
\ArrowLine(0,0)(19,25)\Text(10,12)[lt]{$\scriptscriptstyle e(k_1)$}
\ArrowLine(19,25)(0,50)\Text(10,38)[lb]{$\scriptscriptstyle e(k_2)$}
\Line(19,25)(50,25)\Text(34,26)[b]{$\scriptscriptstyle Z(k_3)^\mu$}
\end{picture}
&
ig_z\left
(	\frac12\gamma^\mu p_l
	-\gamma^\mu\sin^2\theta_w
	+\frac12\frac{m_e}{k_3^2}k_3^\mu\gamma^5\right
)
\cr
\begin{picture}(50,50)(0,0)
\ArrowLine(0,0)(19,25)\Text(10,12)[lt]{$\scriptscriptstyle e(k_1)$}
\ArrowLine(19,25)(0,50)\Text(10,38)[lb]{$\scriptscriptstyle e(k_2)$}
\Line(19,25)(50,25)\Text(34,26)[b]{$\scriptscriptstyle \phi_Z(k_3)$}
\end{picture}
&
\frac12g_z\frac{m_e}{M_Z}\gamma^5
\cr
\begin{picture}(50,50)(0,0)
\ArrowLine(0,0)(19,25)\Text(10,12)[lt]{$\scriptscriptstyle e(k_1)$}
\ArrowLine(19,25)(0,50)\Text(10,38)[lb]{$\scriptscriptstyle e(k_2)$}
\Line(19,25)(50,25)\Text(34,26)[b]{$\scriptscriptstyle H(k_3)$}
\end{picture}
&
-\frac i2g_w\frac{m_e}{M_W}
\cr
\begin{picture}(50,50)(0,0)
\ArrowLine(0,0)(19,25)\Text(10,12)[lt]{$\scriptscriptstyle e(k_1)$}
\ArrowLine(19,25)(0,50)\Text(10,38)[lb]{$\scriptscriptstyle \nu_e(k_2)$}
\ArrowLine(19,25)(50,25)\Text(34,26)[b]{$\scriptscriptstyle W(k_3)^\mu$}
\end{picture}
&
-\frac{ig_w}{\sqrt2}V_{11}\left(
	\gamma^\mu p_l
	+\left(m_{\nu_e}p_l-m_e p_r\right)\frac{k_3^\mu}{k_3^2}
\right)
\cr
\begin{picture}(50,50)(0,0)
\ArrowLine(0,0)(19,25)\Text(10,12)[lt]{$\scriptscriptstyle e(k_1)$}
\ArrowLine(19,25)(0,50)\Text(10,38)[lb]{$\scriptscriptstyle \nu_e(k_2)$}
\ArrowLine(19,25)(50,25)\Text(34,26)[b]{$\scriptscriptstyle \phi_W(k_3)$}
\end{picture}
&
\frac i{\sqrt2}\frac{g_w}{M_W}V_{11}\left(m_{\nu_e}p_l-m_ep_r\right)
\cr
\begin{picture}(50,50)(0,0)
\ArrowLine(0,0)(19,25)\Text(10,12)[lt]{$\scriptscriptstyle \nu_e(k_1)$}
\ArrowLine(19,25)(0,50)\Text(10,38)[lb]{$\scriptscriptstyle \nu_e(k_2)$}
\Line(19,25)(50,25)\Text(34,26)[b]{$\scriptscriptstyle Z(k_3)^\mu$}
\end{picture}
&
-\frac i2g_z\left(\gamma^\mu p_l+m_{\nu_e}\gamma^5\frac{k_3^\mu}{k_3^2}\right)
\cr
\begin{picture}(50,50)(0,0)
\ArrowLine(0,0)(19,25)\Text(10,12)[lt]{$\scriptscriptstyle \nu_e(k_1)$}
\ArrowLine(19,25)(0,50)\Text(10,38)[lb]{$\scriptscriptstyle \nu_e(k_2)$}
\Line(19,25)(50,25)\Text(34,26)[b]{$\scriptscriptstyle \phi_Z(k_3)$}
\end{picture}
&
-\frac12 g_w\frac{m_{\nu_e}}{M_W}\gamma^5
\cr
\begin{picture}(50,50)(0,0)
\ArrowLine(0,0)(19,25)\Text(10,12)[lt]{$\scriptscriptstyle \nu_e(k_1)$}
\ArrowLine(19,25)(0,50)\Text(10,38)[lb]{$\scriptscriptstyle \nu_e(k_2)$}
\Line(19,25)(50,25)\Text(34,26)[b]{$\scriptscriptstyle H(k_3)$}
\end{picture}
&
-\frac i2g_w\frac{m_{\nu_e}}{M_W}
\cr
\begin{picture}(50,50)(0,0)
\ArrowLine(19,25)(0,0)\Text(10,12)[lt]{$\scriptscriptstyle \mu(k_1)$}
\ArrowLine(0,50)(19,25)\Text(10,38)[lb]{$\scriptscriptstyle \nu_e(k_2)$}
\ArrowLine(50,25)(19,25)\Text(34,27)[b]{$\scriptscriptstyle W(k_3)^\mu$}
\end{picture}
&
-\frac{ig_w}{\sqrt2}V^\dagger_{21}\left(
	\gamma^\mu p_l
	-\left(m_\mu p_l-m_{\nu_e}p_r\right)\frac{k_3^\mu}{k_3^2}
\right)
\cr
\begin{picture}(50,50)(0,0)
\ArrowLine(19,25)(0,0)\Text(10,12)[lt]{$\scriptscriptstyle \mu(k_1)$}
\ArrowLine(0,50)(19,25)\Text(10,38)[lb]{$\scriptscriptstyle \nu_e(k_2)$}
\ArrowLine(50,25)(19,25)\Text(34,27)[b]{$\scriptscriptstyle \phi_W(k_3)$}
\end{picture}
&
-\frac{i}{\sqrt2}\frac{g_w}{M_W}V^\dagger_{21}\left(
	m_\mu p_l
	-m_{\nu_e}p_r
\right)
\cr
}

\subsection{Quadruple boson couplings among $B$, $W$ and $\phi_W$}
\fmruletab{
\begin{picture}(50,50)(0,0)
\ArrowLine(0,0)(25,25)\Text(12,13)[rb]{$\scriptscriptstyle W(k_1)^\mu$}
\ArrowLine(25,25)(0,50)\Text(12,37)[rt]{$\scriptscriptstyle W(k_2)^\nu$}
\Line(50,0)(25,25)\Text(38,13)[lb]{$\scriptscriptstyle B(k_3)^\sigma$}
\Line(50,50)(25,25)\Text(38,37)[lt]{$\scriptscriptstyle B(k_4)^\tau$}
\end{picture}
&
ig_e^2\left(
	-2g^{\mu\nu}g^{\sigma\tau}
	+g^{\mu\sigma}g^{\nu\tau}
	+g^{\mu\tau}g^{\nu\sigma}
	+2M_W^2g^{\sigma\tau}\frac{k_1^\mu}{k_1^2}\frac{k_2^\nu}{k_2^2}\right)
\cr
\begin{picture}(50,50)(0,0)
\ArrowLine(0,0)(25,25)\Text(12,13)[rb]{$\scriptscriptstyle W(k_1)^\mu$}
\ArrowLine(0,50)(25,25)\Text(12,37)[rt]{$\scriptscriptstyle W(k_2)^\nu$}
\ArrowLine(25,25)(50,0)\Text(38,13)[lb]{$\scriptscriptstyle W(k_3)^\sigma$}
\ArrowLine(25,25)(50,50)\Text(38,37)[lt]{$\scriptscriptstyle W(k_4)^\tau$}
\end{picture}
&
\eqalign{
ig_w^2\left[\vphantom{\frac12}\right.&
	2g^{\mu\nu}g^{\sigma\tau}
	-g^{\mu\sigma}g^{\nu\tau}
	-g^{\mu\tau}g^{\nu\sigma}\cr
	&+\frac12M_W^2\left(
		g^{\nu\tau}\frac{k_1^\mu}{k_1^2}\frac{k_3^\sigma}{k_3^2}
		+g^{\nu\sigma}\frac{k_1^\mu}{k_1^2}\frac{k_4^\tau}{k_4^2}
		+g^{\mu\tau}\frac{k_2^\nu}{k_2^2}\frac{k_3^\sigma}{k_3^2}
		+g^{\mu\sigma}\frac{k_2^\nu}{k_2^2}\frac{k_4^\tau}{k_4^2}
	\right)\cr
	&\left.-\frac12M_W^2M_H^2\frac{k_1^\mu}{k_1^2}\frac{k_2^\nu}{k_2^2}
										\frac{k_3^\sigma}{k_3^2}\frac{k_4^\tau}{k_4^2}
\right]\cr
}
\cr
\begin{picture}(50,50)(0,0)
\ArrowLine(25,25)(0,0)\Text(12,13)[rb]{$\scriptscriptstyle \phi_W(k_1)$}
\ArrowLine(0,50)(25,25)\Text(12,37)[rt]{$\scriptscriptstyle W(k_2)^\mu$}
\Line(50,0)(25,25)\Text(38,13)[lb]{$\scriptscriptstyle B(k_3)^\nu$}
\Line(50,50)(25,25)\Text(38,37)[lt]{$\scriptscriptstyle B(k_4)^\sigma$}
\end{picture}
&
-2ig_e^2M_Wg^{\nu\sigma}\frac{k_2^\mu}{k_2^2}
\cr
\begin{picture}(50,50)(0,0)
\ArrowLine(25,25)(0,0)\Text(12,13)[rb]{$\scriptscriptstyle \phi_W(k_1)$}
\ArrowLine(0,50)(25,25)\Text(12,37)[rt]{$\scriptscriptstyle W(k_2)^\mu$}
\ArrowLine(50,0)(25,25)\Text(38,13)[lb]{$\scriptscriptstyle W(k_3)^\nu$}
\ArrowLine(25,25)(50,50)\Text(38,37)[lt]{$\scriptscriptstyle W(k_4)^\sigma$}
\end{picture}
&
\frac i2g_w^2M_W\left(
	-g^{\nu\sigma}\frac{k_2^\mu}{k_2^2}
	-g^{\mu\sigma}\frac{k_3^\nu}{k_3^2}
	+M_H^2\frac{k_2^\mu}{k_2^2}\frac{k_3^\nu}{k_3^2}
		\frac{k_4^\sigma}{k_4^2}\right)
\cr
\begin{picture}(50,50)(0,0)
\ArrowLine(25,25)(0,0)\Text(12,13)[rb]{$\scriptscriptstyle \phi_W(k_1)$}
\ArrowLine(25,25)(0,50)\Text(12,37)[rt]{$\scriptscriptstyle \phi_W(k_2)$}
\ArrowLine(50,0)(25,25)\Text(38,13)[lb]{$\scriptscriptstyle W(k_3)^\mu$}
\ArrowLine(50,50)(25,25)\Text(38,37)[lt]{$\scriptscriptstyle W(k_4)^\nu$}
\end{picture}
&-\frac i2 g_w^2M_H^2\frac{k_3^\mu}{k_3^2}\frac{k_4^\nu}{k_4^2}
\cr
\begin{picture}(50,50)(0,0)
\ArrowLine(0,0)(25,25)\Text(12,13)[rb]{$\scriptscriptstyle \phi_W(k_1)$}
\ArrowLine(25,25)(0,50)\Text(12,37)[rt]{$\scriptscriptstyle \phi_W(k_2)$}
\Line(50,0)(25,25)\Text(38,13)[lb]{$\scriptscriptstyle B(k_3)^\mu$}
\Line(50,50)(25,25)\Text(38,37)[lt]{$\scriptscriptstyle B(k_4)^\nu$}
\end{picture}
&
2ig_e^2g^{\mu\nu}
\cr
\begin{picture}(50,50)(0,0)
\ArrowLine(0,0)(25,25)\Text(12,13)[rb]{$\scriptscriptstyle \phi_W(k_1)$}
\ArrowLine(25,25)(0,50)\Text(12,37)[rt]{$\scriptscriptstyle \phi_W(k_2)$}
\ArrowLine(50,0)(25,25)\Text(38,13)[lb]{$\scriptscriptstyle W(k_3)^\mu$}
\ArrowLine(25,25)(50,50)\Text(38,37)[lt]{$\scriptscriptstyle W(k_4)^\nu$}
\end{picture}
&
\frac i2g_w^2\left(
	g^{\mu\nu}-M_H^2\frac{k_3^\mu}{k_3^2}\frac{k_4^\nu}{k_4^2}\right)
\cr
\begin{picture}(50,50)(0,0)
\ArrowLine(0,0)(25,25)\Text(12,13)[rb]{$\scriptscriptstyle \phi_W(k_1)$}
\ArrowLine(25,25)(0,50)\Text(12,37)[rt]{$\scriptscriptstyle \phi_W(k_2)$}
\ArrowLine(25,25)(50,0)\Text(38,13)[lb]{$\scriptscriptstyle \phi_W(k_3)$}
\ArrowLine(50,50)(25,25)\Text(38,37)[lt]{$\scriptscriptstyle W(k_4)^\mu$}
\end{picture}
&
\frac i2g_w^2\frac{M_H^2}{M_W}\frac{k_4^\mu}{k_4^2}
\cr
\begin{picture}(50,50)(0,0)
\ArrowLine(0,0)(25,25)\Text(12,13)[rb]{$\scriptscriptstyle \phi_W(k_1)$}
\ArrowLine(0,50)(25,25)\Text(12,37)[rt]{$\scriptscriptstyle \phi_W(k_2)$}
\ArrowLine(25,25)(50,0)\Text(38,13)[lb]{$\scriptscriptstyle \phi_W(k_3)$}
\ArrowLine(25,25)(50,50)\Text(38,37)[lt]{$\scriptscriptstyle \phi_W(k_4)$}
\end{picture}
&
-\frac i2g_w^2\frac{M_H^2}{M_W^2}
\cr
}

\subsection{Quadruple boson couplings with $Z$, and without $\phi_Z$ or $H$}
\fmruletab{
\begin{picture}(50,50)(0,0)
\Line(25,25)(0,0)\Text(12,13)[rb]{$\scriptscriptstyle Z(k_1)^\mu$}
\ArrowLine(0,50)(25,25)\Text(12,37)[rt]{$\scriptscriptstyle W(k_2)^\nu$}
\ArrowLine(25,25)(50,0)\Text(38,13)[lb]{$\scriptscriptstyle W(k_3)^\sigma$}
\Line(25,25)(50,50)\Text(38,37)[lt]{$\scriptscriptstyle B(k_4)^\tau$}
\end{picture}
&
\eqalign{
ig_eg_w\cos\theta_w\left[
	-2g^{\mu\tau}g^{\nu\sigma}
	+g^{\mu\nu}g^{\sigma\tau}
	+g^{\mu\sigma}g^{\nu\tau}+M_Z^2\left(
		\frac12g^{\sigma\tau}\frac{k_1^\mu}{k_1^2}\frac{k_2^\nu}{k_2^2}\right.\right.\cr
	\left.\left.-\frac12g^{\nu\tau}\frac{k_1^\mu}{k_1^2}\frac{k_3^\sigma}{k_3^2}
		+g^{\mu\tau}(2\cos\theta_w^2-1)\frac{k_2^\nu}{k_2^2}
																	\frac{k_3^\sigma}{k_3^2}
	\right)
\right]\cr
}
\cr
\begin{picture}(50,50)(0,0)
\Line(25,25)(0,0)\Text(12,13)[rb]{$\scriptscriptstyle Z(k_1)^\mu$}
\ArrowLine(25,25)(0,50)\Text(12,37)[rt]{$\scriptscriptstyle \phi_W(k_2)$}
\ArrowLine(50,0)(25,25)\Text(38,13)[lb]{$\scriptscriptstyle W(k_3)^\nu$}
\Line(25,25)(50,50)\Text(38,37)[lt]{$\scriptscriptstyle B(k_4)^\sigma$}
\end{picture}
&
ig_eg_zM_W\left(
	\frac12g^{\nu\sigma}\frac{k_1^\mu}{k_1^2}
	+g^{\mu\sigma}(1-2\cos^2\theta_w)\frac{k_3^\nu}{k_3^2}
\right)
\cr
\begin{picture}(50,50)(0,0)
\Line(25,25)(0,0)\Text(12,13)[rb]{$\scriptscriptstyle Z(k_1)^\mu$}
\ArrowLine(0,50)(25,25)\Text(12,37)[rt]{$\scriptscriptstyle \phi_W(k_2)$}
\ArrowLine(25,25)(50,0)\Text(38,13)[lb]{$\scriptscriptstyle \phi_W(k_3)$}
\Line(25,25)(50,50)\Text(38,37)[lt]{$\scriptscriptstyle B(k_4)^\nu$}
\end{picture}
&
ig_eg_z(2\cos^2\theta_w-1)g^{\mu\nu}
\cr
\begin{picture}(50,50)(0,0)
\Line(25,25)(0,0)\Text(12,13)[rb]{$\scriptscriptstyle Z(k_1)^\mu$}
\Line(0,50)(25,25)\Text(12,37)[rt]{$\scriptscriptstyle Z(k_2)^\nu$}
\ArrowLine(50,0)(25,25)\Text(38,13)[lb]{$\scriptscriptstyle W(k_3)^\sigma$}
\ArrowLine(25,25)(50,50)\Text(38,37)[lt]{$\scriptscriptstyle W(k_4)^\tau$}
\end{picture}
&
\eqalign{
-ig_w^2\left[\vphantom{\frac12}\right.
	&\cos^2\theta_w\left(
		2g^{\mu\nu}g^{\sigma\tau}
		-g^{\mu\sigma}g^{\nu\tau}
		-g^{\mu\tau}g^{\nu\sigma}
	\right)\cr
	&+\frac12M_Z^2\sin^2\theta_w\left(
		\frac1{\sin^2\theta_w}g^{\sigma\tau}\frac{k_1^\mu}{k_1^2}
																		\frac{k_2^\nu}{k_2^2}
		+g^{\nu\tau}\frac{k_1^\mu}{k_1^2}\frac{k_3^\sigma}{k_3^2}
		-g^{\nu\sigma}\frac{k_1^\mu}{k_1^2}\frac{k_4^\tau}{k_4^2}\right.\cr
		&+g^{\mu\tau}\frac{k_2^\nu}{k_2^2}\frac{k_3^\sigma}{k_3^2}
		-g^{\mu\sigma}\frac{k_2^\nu}{k_2^2}\frac{k_4^\tau}{k_4^2}
		+\left(4\cos^2\theta_w-\frac1{\sin^2\theta_w}\right)g^{\mu\nu}
										\frac{k_3^\sigma}{k_3^2}\frac{k_4^\tau}{k_4^2}\cr
		&\left.\left.-\frac{M_H^2}{2\sin^2\theta_w}\frac{k_1^\mu}{k_1^2}
				\frac{k_2^\nu}{k_2^2}\frac{k_3^\sigma}{k_3^2}\frac{k_4^\tau}{k_4^2}
	\right)
\right]\cr
}
\cr
\begin{picture}(50,50)(0,0)
\Line(25,25)(0,0)\Text(12,13)[rb]{$\scriptscriptstyle Z(k_1)^\mu$}
\Line(0,50)(25,25)\Text(12,37)[rt]{$\scriptscriptstyle Z(k_2)^\nu$}
\ArrowLine(25,25)(50,0)\Text(38,13)[lb]{$\scriptscriptstyle \phi_W(k_3)$}
\ArrowLine(50,50)(25,25)\Text(38,37)[lt]{$\scriptscriptstyle W(k_4)^\sigma$}
\end{picture}
&
\eqalign{
-\frac i2\frac{g_e^2M_W}{\cos^2\theta_w}\left[\vphantom{\frac12}\right.
	g^{\nu\sigma}\frac{k_1^\mu}{k_1^2}
	+g^{\mu\sigma}\frac{k_2^\nu}{k_2^2}
	+\left(\frac1{\sin^2\theta_w}-4\cos^2\theta_w\right)g^{\mu\nu}
																	\frac{k_4^\sigma}{k_4^2}\cr
	\left.+\frac1{2\sin^2\theta_w}M_H^2\frac{k_1^\mu}{k_1^2}
											\frac{k_2^\nu}{k_2^2}\frac{k_4^\sigma}{k_4^2}
\right]\cr
}
\cr
\begin{picture}(50,50)(0,0)
\Line(25,25)(0,0)\Text(12,13)[rb]{$\scriptscriptstyle Z(k_1)^\mu$}
\Line(25,25)(0,50)\Text(12,37)[rt]{$\scriptscriptstyle Z(k_2)^\nu$}
\ArrowLine(50,0)(25,25)\Text(38,13)[lb]{$\scriptscriptstyle \phi_W(k_3)$}
\ArrowLine(25,25)(50,50)\Text(38,37)[lt]{$\scriptscriptstyle \phi_W(k_4)$}
\end{picture}
&
ig_z^2\left(
	\left(\frac12-2\cos^2\theta_w\sin^2\theta_w\right)g^{\mu\nu}
	+\frac14M_H^2\frac{k_1^\mu}{k_1^2}\frac{k_2^\nu}{k_2^2}
\right)
\cr
\begin{picture}(50,50)(0,0)
\Line(25,25)(0,0)\Text(12,13)[rb]{$\scriptscriptstyle Z(k_1)^\mu$}
\Line(25,25)(0,50)\Text(12,37)[rt]{$\scriptscriptstyle Z(k_2)^\nu$}
\Line(50,0)(25,25)\Text(38,13)[lb]{$\scriptscriptstyle Z(k_3)^\sigma$}
\Line(25,25)(50,50)\Text(38,37)[lt]{$\scriptscriptstyle Z(k_4)^\tau$}
\end{picture}
&
\eqalign{
	-\frac i2g_z^2M_Z^2\left(\vphantom{\frac12}\right.&
		g^{\mu\nu}\frac{k_3^\sigma}{k_3^2}\frac{k_4^\tau}{k_4^2}
		+g^{\mu\sigma}\frac{k_2^\nu}{k_2^2}\frac{k_4^\tau}{k_4^2}
		+g^{\mu\tau}\frac{k_2^\nu}{k_2^2}\frac{k_3^\sigma}{k_3^2}
		+g^{\nu\sigma}\frac{k_1^\mu}{k_1^2}\frac{k_4^\tau}{k_4^2}\cr
		&\left.+g^{\nu\tau}\frac{k_1^\mu}{k_1^2}\frac{k_3^\sigma}{k_3^2}
		+g^{\sigma\tau}\frac{k_1^\mu}{k_1^2}\frac{k_2^\nu}{k_2^2}
		+\frac32M_H^2\frac{k_1^\mu}{k_1^2}\frac{k_2^\nu}{k_2^2}
											\frac{k_3^\sigma}{k_3^2}\frac{k_4^\tau}{k_4^2}
	\right)
}
\cr
}

\subsection{Quadruple boson couplings with one $\phi_Z$ and no $H$}
\fmruletab{
\begin{picture}(50,50)(0,0)
\Line(25,25)(0,0)\Text(12,13)[rb]{$\scriptscriptstyle \phi_Z(k_1)$}
\ArrowLine(0,50)(25,25)\Text(12,37)[rt]{$\scriptscriptstyle W(k_2)^\mu$}
\ArrowLine(25,25)(50,0)\Text(38,13)[lb]{$\scriptscriptstyle W(k_3)^\nu$}
\Line(25,25)(50,50)\Text(38,37)[lt]{$\scriptscriptstyle B(k_4)^\sigma$}
\end{picture}
&
\frac12g_eg_wM_W\left(
	g^{\nu\sigma}\frac{k_2^\mu}{k_2^2}
	-g^{\mu\sigma}\frac{k_3^\nu}{k_3^2}
\right)
\cr
\begin{picture}(50,50)(0,0)
\Line(25,25)(0,0)\Text(12,13)[rb]{$\scriptscriptstyle \phi_Z(k_1)$}
\ArrowLine(25,25)(0,50)\Text(12,37)[rt]{$\scriptscriptstyle \phi_W(k_2)$}
\ArrowLine(50,0)(25,25)\Text(38,13)[lb]{$\scriptscriptstyle W(k_3)^\mu$}
\Line(25,25)(50,50)\Text(38,37)[lt]{$\scriptscriptstyle B(k_4)^\nu$}
\end{picture}
&
\frac12g_eg_wg^{\mu\nu}
\cr
\begin{picture}(50,50)(0,0)
\Line(25,25)(0,0)\Text(12,13)[rb]{$\scriptscriptstyle \phi_Z(k_1)$}
\Line(25,25)(0,50)\Text(12,37)[rt]{$\scriptscriptstyle Z(k_2)^\mu$}
\ArrowLine(50,0)(25,25)\Text(38,13)[lb]{$\scriptscriptstyle W(k_3)^\nu$}
\ArrowLine(25,25)(50,50)\Text(38,37)[lt]{$\scriptscriptstyle W(k_4)^\sigma$}
\end{picture}
&
\frac12g_e^2M_Z\left(
	-\frac{g^{\nu\sigma}}{\sin^2\theta_w}\frac{k_2^\mu}{k_2^2}
	-g^{\mu\sigma}\frac{k_3^\nu}{k_3^2}
	+g^{\mu\nu}\frac{k_4^\sigma}{k_4^2}
	+\frac12\frac{M_H^2}{\sin^2\theta_w}\frac{k_2^\mu}{k_2^2}
			\frac{k_3^\nu}{k_3^2}\frac{k_4^\sigma}{k_4^2}
\right)
\cr
\begin{picture}(50,50)(0,0)
\Line(25,25)(0,0)\Text(12,13)[rb]{$\scriptscriptstyle \phi_Z(k_1)$}
\Line(25,25)(0,50)\Text(12,37)[rt]{$\scriptscriptstyle Z(k_2)^\mu$}
\ArrowLine(25,25)(50,0)\Text(38,13)[lb]{$\scriptscriptstyle \phi_W(k_3)$}
\ArrowLine(50,50)(25,25)\Text(38,37)[lt]{$\scriptscriptstyle W(k_4)^\nu$}
\end{picture}
&
-g_e^2\left(\frac1{2\cos\theta_w}g^{\mu\nu}+\frac{M_H^2}{4\sin^2\theta_w\cos\theta_w}\frac{k_2^\mu}{k_2^2}\frac{k_4^\nu}{k_4^2}\right)
\cr
\begin{picture}(50,50)(0,0)
\Line(25,25)(0,0)\Text(12,13)[rb]{$\scriptscriptstyle \phi_Z(k_1)$}
\Line(25,25)(0,50)\Text(12,37)[rt]{$\scriptscriptstyle Z(k_2)^\mu$}
\ArrowLine(50,0)(25,25)\Text(38,13)[lb]{$\scriptscriptstyle \phi_W(k_3)$}
\ArrowLine(25,25)(50,50)\Text(38,37)[lt]{$\scriptscriptstyle \phi_W(k_4)$}
\end{picture}
&
\frac14g_z^2\frac{M_H^2}{M_Z}\frac{k_2^\mu}{k_2^2}
\cr
\begin{picture}(50,50)(0,0)
\Line(25,25)(0,0)\Text(12,13)[rb]{$\scriptscriptstyle \phi_Z(k_1)$}
\Line(25,25)(0,50)\Text(12,37)[rt]{$\scriptscriptstyle Z(k_2)^\mu$}
\Line(50,0)(25,25)\Text(38,13)[lb]{$\scriptscriptstyle Z(k_3)^\nu$}
\Line(25,25)(50,50)\Text(38,37)[lt]{$\scriptscriptstyle Z(k_4)^\sigma$}
\end{picture}
&
-\frac12g_z^2M_Z\left(
	g^{\nu\sigma}\frac{k_2^\mu}{k_2^2}
	+g^{\mu\sigma}\frac{k_3^\nu}{k_3^2}
	+g^{\mu\nu}\frac{k_4^\sigma}{k_4^2}
	+\frac32M_H^2\frac{k_2^\mu}{k_2^2}\frac{k_3^\nu}{k_3^2}
																		\frac{k_4^\sigma}{k_4^2}
\right)
\cr
}

\subsection{Quadruple boson couplings with multiple $\phi_Z$ and no $H$}
\fmruletab{
\begin{picture}(50,50)(0,0)
\Line(25,25)(0,0)\Text(12,13)[rb]{$\scriptscriptstyle \phi_Z(k_1)$}
\Line(25,25)(0,50)\Text(12,37)[rt]{$\scriptscriptstyle \phi_Z(k_2)$}
\ArrowLine(50,0)(25,25)\Text(38,13)[lb]{$\scriptscriptstyle W(k_3)^\mu$}
\ArrowLine(25,25)(50,50)\Text(38,37)[lt]{$\scriptscriptstyle W(k_4)^\nu$}
\end{picture}
&
ig_w^2\left(\frac12g^{\mu\nu}-\frac14M_H^2\frac{k_3^\mu}{k_3^2}
																\frac{k_4^\nu}{k_4^2}\right)
\cr
\begin{picture}(50,50)(0,0)
\Line(25,25)(0,0)\Text(12,13)[rb]{$\scriptscriptstyle \phi_Z(k_1)$}
\Line(25,25)(0,50)\Text(12,37)[rt]{$\scriptscriptstyle \phi_Z(k_2)$}
\ArrowLine(25,25)(50,0)\Text(38,13)[lb]{$\scriptscriptstyle \phi_W(k_3)$}
\ArrowLine(50,50)(25,25)\Text(38,37)[lt]{$\scriptscriptstyle W(k_4)^\mu$}
\end{picture}
&
\frac i4g_w^2\frac{M_H^2}{M_W}\frac{k_4^\mu}{k_4^2}
\cr
\begin{picture}(50,50)(0,0)
\Line(25,25)(0,0)\Text(12,13)[rb]{$\scriptscriptstyle \phi_Z(k_1)$}
\Line(25,25)(0,50)\Text(12,37)[rt]{$\scriptscriptstyle \phi_Z(k_2)$}
\ArrowLine(50,0)(25,25)\Text(38,13)[lb]{$\scriptscriptstyle \phi_W(k_3)$}
\ArrowLine(25,25)(50,50)\Text(38,37)[lt]{$\scriptscriptstyle \phi_W(k_4)$}
\end{picture}
&
-\frac i4g_w^2\frac{M_H^2}{M_W^2}
\cr
\begin{picture}(50,50)(0,0)
\Line(25,25)(0,0)\Text(12,13)[rb]{$\scriptscriptstyle \phi_Z(k_1)$}
\Line(25,25)(0,50)\Text(12,37)[rt]{$\scriptscriptstyle \phi_Z(k_2)$}
\Line(50,0)(25,25)\Text(38,13)[lb]{$\scriptscriptstyle Z(k_3)^\mu$}
\Line(25,25)(50,50)\Text(38,37)[lt]{$\scriptscriptstyle Z(k_4)^\nu$}
\end{picture}
&
\frac i2g_z^2\left(
	g^{\mu\nu}
	+\frac32M_H^2\frac{k_3^\mu}{k_3^2}\frac{k_4^\nu}{k_4^2}
\right)
\cr
\begin{picture}(50,50)(0,0)
\Line(25,25)(0,0)\Text(12,13)[rb]{$\scriptscriptstyle \phi_Z(k_1)$}
\Line(25,25)(0,50)\Text(12,37)[rt]{$\scriptscriptstyle \phi_Z(k_2)$}
\Line(50,0)(25,25)\Text(38,13)[lb]{$\scriptscriptstyle \phi_Z(k_3)$}
\Line(50,50)(25,25)\Text(38,37)[lt]{$\scriptscriptstyle Z(k_4)^\mu$}
\end{picture}
&
\frac34g_z^2\frac{M_H^2}{M_Z}\frac{k_4^\mu}{k_4^2}
\cr
\begin{picture}(50,50)(0,0)
\Line(25,25)(0,0)\Text(12,13)[rb]{$\scriptscriptstyle \phi_Z(k_1)$}
\Line(25,25)(0,50)\Text(12,37)[rt]{$\scriptscriptstyle \phi_Z(k_2)$}
\Line(50,0)(25,25)\Text(38,13)[lb]{$\scriptscriptstyle \phi_Z(k_3)$}
\Line(50,50)(25,25)\Text(38,37)[lt]{$\scriptscriptstyle \phi_Z(k_4)$}
\end{picture}
&
-\frac34ig_z^2\frac{M_H^2}{M_Z^2}
\cr
}
\subsection{Quadruple boson couplings with one $H$}
\fmruletab{
\begin{picture}(50,50)(0,0)
\Line(25,25)(0,0)\Text(12,13)[rb]{$\scriptscriptstyle H(k_1)$}
\ArrowLine(0,50)(25,25)\Text(12,37)[rt]{$\scriptscriptstyle W(k_2)^\mu$}
\ArrowLine(25,25)(50,0)\Text(38,13)[lb]{$\scriptscriptstyle W(k_3)^\nu$}
\Line(50,50)(25,25)\Text(38,37)[lt]{$\scriptscriptstyle B(k_4)^\sigma$}
\end{picture}
&
-\frac i2g_eg_wM_W\left(
	g^{\nu\sigma}\frac{k_2^\mu}{k_2^2}
	+g^{\mu\sigma}\frac{k_3^\nu}{k_3^2}
\right)
\cr
\begin{picture}(50,50)(0,0)
\Line(25,25)(0,0)\Text(12,13)[rb]{$\scriptscriptstyle H(k_1)$}
\ArrowLine(25,25)(0,50)\Text(12,37)[rt]{$\scriptscriptstyle \phi_W(k_2)$}
\ArrowLine(50,0)(25,25)\Text(38,13)[lb]{$\scriptscriptstyle W(k_3)^\mu$}
\Line(50,50)(25,25)\Text(38,37)[lt]{$\scriptscriptstyle B(k_4)^\nu$}
\end{picture}
&
\frac i2g_eg_wg^{\mu\nu}
\cr
\begin{picture}(50,50)(0,0)
\Line(25,25)(0,0)\Text(12,13)[rb]{$\scriptscriptstyle H(k_1)$}
\Line(25,25)(0,50)\Text(12,37)[rt]{$\scriptscriptstyle Z(k_2)^\mu$}
\ArrowLine(50,0)(25,25)\Text(38,13)[lb]{$\scriptscriptstyle W(k_3)^\nu$}
\ArrowLine(25,25)(50,50)\Text(38,37)[lt]{$\scriptscriptstyle W(k_4)^\sigma$}
\end{picture}
&
\frac i2g_e^2M_Z\left(
	g^{\mu\sigma}\frac{k_3^\nu}{k_3^2}
	+g^{\mu\nu}\frac{k_4^\sigma}{k_4^2}
\right)
\cr
\begin{picture}(50,50)(0,0)
\Line(25,25)(0,0)\Text(12,13)[rb]{$\scriptscriptstyle H(k_1)$}
\Line(25,25)(0,50)\Text(12,37)[rt]{$\scriptscriptstyle Z(k_2)^\mu$}
\ArrowLine(25,25)(50,0)\Text(38,13)[lb]{$\scriptscriptstyle \phi_W(k_3)$}
\ArrowLine(50,50)(25,25)\Text(38,37)[lt]{$\scriptscriptstyle W(k_4)^\nu$}
\end{picture}
&
-\frac i2\frac{g_e^2}{\cos\theta_w}g^{\mu\nu}
\cr
}

\subsection{Quadruple boson couplings with multiple $H$}
\fmruletab{
\begin{picture}(50,50)(0,0)
\Line(25,25)(0,0)\Text(12,13)[rb]{$\scriptscriptstyle H(k_1)$}
\Line(25,25)(0,50)\Text(12,37)[rt]{$\scriptscriptstyle H(k_2)$}
\ArrowLine(50,0)(25,25)\Text(38,13)[lb]{$\scriptscriptstyle W(k_3)^\mu$}
\ArrowLine(25,25)(50,50)\Text(38,37)[lt]{$\scriptscriptstyle W(k_4)^\nu$}
\end{picture}
&
ig_w^2\left(\frac12g^{\mu\nu}
	-\frac14M_H^2\frac{k_3^\mu}{k_3^2}\frac{k_4^\nu}{k_4^2}\right)
\cr
\begin{picture}(50,50)(0,0)
\Line(25,25)(0,0)\Text(12,13)[rb]{$\scriptscriptstyle H(k_1)$}
\Line(25,25)(0,50)\Text(12,37)[rt]{$\scriptscriptstyle H(k_2)$}
\ArrowLine(25,25)(50,0)\Text(38,13)[lb]{$\scriptscriptstyle \phi_W(k_3)$}
\ArrowLine(50,50)(25,25)\Text(38,37)[lt]{$\scriptscriptstyle W(k_4)^\mu$}
\end{picture}
&
\frac i4g_w^2\frac{M_H^2}{M_W}\frac{k_4^\mu}{k_4^2}
\cr
\begin{picture}(50,50)(0,0)
\Line(25,25)(0,0)\Text(12,13)[rb]{$\scriptscriptstyle H(k_1)$}
\Line(25,25)(0,50)\Text(12,37)[rt]{$\scriptscriptstyle H(k_2)$}
\ArrowLine(50,0)(25,25)\Text(38,13)[lb]{$\scriptscriptstyle \phi_W(k_3)$}
\ArrowLine(25,25)(50,50)\Text(38,37)[lt]{$\scriptscriptstyle \phi_W(k_4)$}
\end{picture}
&
-\frac i4g_w^2\frac{M_H^2}{M_W^2}
\cr
\begin{picture}(50,50)(0,0)
\Line(25,25)(0,0)\Text(12,13)[rb]{$\scriptscriptstyle H(k_1)$}
\Line(25,25)(0,50)\Text(12,37)[rt]{$\scriptscriptstyle H(k_2)$}
\Line(25,25)(50,0)\Text(38,13)[lb]{$\scriptscriptstyle Z(k_3)^\mu$}
\Line(50,50)(25,25)\Text(38,37)[lt]{$\scriptscriptstyle Z(k_4)^\nu$}
\end{picture}
&
\frac i2g_z^2\left(
	g^{\mu\nu}
	+\frac12M_H^2\frac{k_3^\mu}{k_3^2}\frac{k_4^\nu}{k_4^2}
\right)
\cr
\begin{picture}(50,50)(0,0)
\Line(25,25)(0,0)\Text(12,13)[rb]{$\scriptscriptstyle H(k_1)$}
\Line(25,25)(0,50)\Text(12,37)[rt]{$\scriptscriptstyle H(k_2)$}
\Line(50,0)(25,25)\Text(38,13)[lb]{$\scriptscriptstyle \phi_Z(k_3)$}
\Line(25,25)(50,50)\Text(38,37)[lt]{$\scriptscriptstyle Z(k_4)^\mu$}
\end{picture}
&
\frac14g_z^2\frac{M_H^2}{M_Z}\frac{k_4^\mu}{k_4^2}
\cr
\begin{picture}(50,50)(0,0)
\Line(25,25)(0,0)\Text(12,13)[rb]{$\scriptscriptstyle H(k_1)$}
\Line(25,25)(0,50)\Text(12,37)[rt]{$\scriptscriptstyle H(k_2)$}
\Line(50,0)(25,25)\Text(38,13)[lb]{$\scriptscriptstyle \phi_Z(k_3)$}
\Line(25,25)(50,50)\Text(38,37)[lt]{$\scriptscriptstyle \phi_Z(k_4)$}
\end{picture}
&
-\frac i4g_w^2\frac{M_H^2}{M_W^2}
\cr
\begin{picture}(50,50)(0,0)
\Line(25,25)(0,0)\Text(12,13)[rb]{$\scriptscriptstyle H(k_1)$}
\Line(25,25)(0,50)\Text(12,37)[rt]{$\scriptscriptstyle H(k_2)$}
\Line(50,0)(25,25)\Text(38,13)[lb]{$\scriptscriptstyle H(k_3)$}
\Line(25,25)(50,50)\Text(38,37)[lt]{$\scriptscriptstyle H(k_4)$}
\end{picture}
&
-\frac34ig_w^2\frac{M_H^2}{M_W^2}
\cr
}

\section{(Un)physical Particles}
The $\phi_W$ and $\phi_Z$ fields are unphysical. This means that they
cannot be external lines in a Feynman graph. The pole at $k^2=0$ that
occurs in their propagators is canceled by the poles in the interaction
vertices that the $W$ and $Z$ particles have. The consequence is that
these particles cannot travel over macroscopic distances. As an example,
we show how this cancellation arrises for one particular case.
Consider the combination
\beq
\Mel
\quad
=
\quad
\vcenter{\hsize=80pt\noindent
\begin{picture}(80,50)(0,0)
\ArrowLine(0,0)(19,25)\Text(10,12)[lt]{$\scriptscriptstyle e(k_1)$}
\ArrowLine(19,25)(0,50)\Text(10,38)[lb]{$\scriptscriptstyle \nu_e(k_2)$}
\ArrowLine(19,25)(61,25)\Text(40,27)[b]{$\scriptscriptstyle W(q)$}
\ArrowLine(80,0)(61,25)\Text(70,12)[rt]{$\scriptscriptstyle \nu_e(k_3)$}
\ArrowLine(61,25)(80,50)\Text(70,38)[rb]{$\scriptscriptstyle e(k_4)$}
\end{picture}
}
\quad
+
\quad
\vcenter{\hsize=80pt\noindent
\begin{picture}(80,50)(0,0)
\ArrowLine(0,0)(19,25)\Text(10,12)[lt]{$\scriptscriptstyle e(k_1)$}
\ArrowLine(19,25)(0,50)\Text(10,38)[lb]{$\scriptscriptstyle \nu_e(k_2)$}
\ArrowLine(19,25)(61,25)\Text(40,27)[b]{$\scriptscriptstyle \phi_W(q)$}
\ArrowLine(80,0)(61,25)\Text(70,12)[rt]{$\scriptscriptstyle \nu_e(k_3)$}
\ArrowLine(61,25)(80,50)\Text(70,38)[rb]{$\scriptscriptstyle e(k_4)$}
\end{picture}
}.
\eeq
We do not assume anything about the external lines here, so that
our conclusions also apply if all lines in the above graphs are internal
lines of some bigger graph. For $\Mel$ we find
\beq
\eqalign{
\Mel&=\frac{ig_w^2}2
		\left[p_r\left(\gamma^\mu-\frac{m_e}{q^2}q^\mu\right)\right]_1
		\frac{
			g_{\mu\nu}
			-\frac{q_\mu n_\nu+q_\nu n_\mu}{q\cdot n}
			+q_\mu q_\nu\frac{n^2}{(q\cdot n)^2}
		}{q^2-M_W^2+i\epsilon}\cr
		&\hskip 7cm
			\left[\left(\gamma^\nu-\frac{m_e}{q^2}q^\nu\right)p_l\right]_2\cr
	&\qquad-\frac{ig_w^2}2\frac{m_e^2}{M_W^2}\left[p_r\right]_1\frac1{q^2}
		\left[p_l\right]_2.
}
\eeq
Here, we have made the approximation that the neutrino's are massless and
consequently the mixing matrix~$V$ can be taken diagonal. This is just
for brevity and does not change much in the proof below.
The $[\cdots]_{1,2}$ are used to distinguish matrices in spinor space for the
two different spin lines. Working out the brackets for the spin lines, we
find
\beq
\label{eq:matel}
\eqalign{
\Mel&=\frac{ig_w^2}2
      \left[p_r\gamma^\mu\right]_1
      \frac{
         g_{\mu\nu}
         -\frac{q_\mu n_\nu+q_\nu n_\mu}{q\cdot n}
         +q_\mu q_\nu\frac{n^2}{(q\cdot n)^2}
      }{q^2-M_W^2+i\epsilon}
			\left[\gamma^\nu p_l\right]_2\cr
	&\qquad+\frac{ig_w^2m_e}2\left[p_r\gamma^\mu\right]_1
		\frac{\frac{n_\mu}{q\cdot n}-q_\mu\frac{n^2}{(q\cdot n)^2}}{
			q^2-M_W^2+i\epsilon}
			\left[p_l\right]_2\cr
	&\qquad+\frac{ig_w^2m_e}2\left[p_r\right]_1
		\frac{\frac{n_\nu}{q\cdot n}-q_\nu\frac{n^2}{(q\cdot n)^2}}{
			q^2-M_W^2+i\epsilon}
			\left[\gamma^\nu p_l\right]_2\cr
	&\qquad-\frac{ig_w^2}2\frac{m_e^2}{q^2}\left[p_r\right]_1
		\frac{1-\frac{q^2n^2}{(q\cdot n)^2}}{q^2-M_W^2+i\epsilon}
		\left[p_l\right]_2\cr
	&\qquad-\frac{ig_w^2}2\frac{m_e^2}{M_W^2}\left[p_r\right]_1\frac1{q^2}
		\left[p_l\right]_2.\cr
}
\eeq
Using the identity
\beq
\frac1{q^2}\frac1{q^2-M_W^2+i\epsilon}
	=\frac1{M_W^2}\frac1{q^2-M_W^2+i\epsilon}
	-\frac1{M_W^2}\frac1{q^2},
\eeq
we see that in equation~\ref{eq:matel} no pole remains at $q^2=0$.

The general property that we need so that this always works out is
that the combination
\beq
\frac{\partial\L_{\textscr{interaction}}}{\partial W_\mu}
	\Delta^W(q)_{\mu\nu}
	\frac{\partial\L_{\textscr{interaction}}}{\partial W^*_\nu}
+\frac{\partial\L_{\textscr{interaction}}}{\partial \phi_W}
	\Delta^{\phi_W}(q)
	\frac{\partial\L_{\textscr{interaction}}}{\partial \phi_W^*}
\eeq
has no pole at $q^2=0$. This property can be checked to hold.
In the same way, it can also be shown that $\phi_Z$ is not a physical
particle.

\section{Outgoing Massive Vector Bosons}
If a massive vector boson is produced in a process, strictly speaking
this cannot be an asymptotic state, and one should take the decay of this
particle into account. However, not doing so may be a rather accurate
approximation. In this section we consider what the r\^ole of the
$\phi_W$~field is. We consider a particular
decay mode of the top quark, namely~$t\to b+\bar b+c$. We compare the result
that can be obtained from the full tree-level matrix element to the
result that we get if we use the $W$~boson as an on-shell particle and to
the result that we get if we ignore the $\phi_W$~field. Notice that the
$\phi_W$ contribution is itself independent of the gauge vector~$n$, and
might therefore be overlooked.
If we consider the $W$~boson as an on-shell particle we find the decay width
\beq
\Gamma_{\textscr{on-shell $W$}}=
	\Gamma_{t\to b+W^+}\frac{\Gamma_{W^+\to\bar b c}}{\Gamma_W}
\eeq
The full tree-level matrix element is given by
\vadjust{\vskip-15pt}
\beq
\Mel\quad=\quad\vcenter{\hsize=75pt\noindent
\begin{picture}(75,75)(0,0)
\ArrowLine(0,37.5)(30,37.5)\Text(15,40)[b]{$\scriptscriptstyle t(p)$}
\ArrowLine(30,37.5)(52.5,60)\Text(42,49)[lt]{$\scriptscriptstyle b(q_1)$}
\ArrowLine(60,15)(30,37.5)\Text(45,27)[lb]{$\scriptscriptstyle W(k)$}
\ArrowLine(75,0)(60,15)\Text(68,8)[lb]{$\scriptscriptstyle \bar b(q_3)$}
\ArrowLine(60,15)(75,30)\Text(68,22)[lt]{$\scriptscriptstyle c(q_2)$}
\end{picture}
}\quad+\quad\vcenter{\hsize=75pt\noindent
\begin{picture}(75,75)(0,0)
\ArrowLine(0,37.5)(30,37.5)\Text(15,40)[b]{$\scriptscriptstyle t(p)$}
\ArrowLine(30,37.5)(52.5,60)\Text(42,49)[lt]{$\scriptscriptstyle b(q_1)$}
\ArrowLine(60,15)(30,37.5)\Text(45,27)[lb]{$\scriptscriptstyle \phi_W(k)$}
\ArrowLine(75,0)(60,15)\Text(68,8)[lb]{$\scriptscriptstyle \bar b(q_3)$}
\ArrowLine(60,15)(75,30)\Text(68,22)[lt]{$\scriptscriptstyle c(q_2)$}
\end{picture}
}\hskip.5cm.
\eeq
We find the following relative errors.
\beq
\eqalign{
\frac{\Gamma_{\textscr{on shell $W$}}-\Gamma_{\textscr{both graphs}}}{
		\Gamma_{\textscr{on shell $W$}}}
	&=\frac{\Gamma_W}{\pi M_W}\left(
		\frac{6M_W^4}{m_t^4+m_t^2M_W^2-2M_W^4}
					\log\left(\frac{m_t^2-M_W^2}{M_W^2}\right)\right.\cr
		&\hskip3cm\left.
			+\frac{m_t^6+3m_t^4M_W^2-6m_t^2M_W^4}{m_t^6-3m_t^2M_W^4+2M_W^6}
		\right)\cr
	&\sim 0.016;
		\cr
\frac{\Gamma_{\textscr{without $\phi_W$}}-\Gamma_{\textscr{both graphs}}}{
		\Gamma_{\textscr{on shell $W$}}}
	&=\frac3{2\pi}\frac{\Gamma_W}{M_W}\frac{m_b^2+m_c^2}{M_W^2}
		\frac{m_t^6}{m_t^6-3m_t^2M_W^4+2M_W^6}\cr
		&\hskip3cm\left(3+\log\left(\frac{m_c+m_b}{m_t}\right)\right)\cr
	&\sim -2\cdot 10^{-5}.
	\cr
}
\eeq
In these expressions we restricted ourselves in both numerator and denominator
to the lowest non-trivial order in $\Gamma_W$, $m_b$ and $m_c$.
What can be learned from this is that because the $\phi_W$ field couples to
the fermions proportional to their mass we expect it not to be important if
either of the fermions the $\phi_W$ couples to has a mass that can be
ignored.

\section{Conclusions}
The electroweak standard model can be considered in the axial gauge.
In this gauge
there are no Fadeev-Popov ghost particles. There are, however, the
unphysical bosons $\phi_W$~and~$\phi_Z$. These bosons cannot appear as
asymptotic states. The $1/k^2$-poles in their propagators cancel against
the $1/k^2$-factors in the vertices of the corresponding physical
particles. The coupling of the fermions to the unphysical fields and to the
$1/k^2$ terms in the vertex factors are proportional to the mass of the
fermions. Consequently, ignoring these masses can be an important
simplification, depending on the amplitude considered.

\end{document}